\documentclass[reprint,aps,pra,amsmath,amssymb,superscriptaddress,letterpaper]{revtex4-2}

\usepackage{iftex}
\ifpdftex\usepackage{cmap}\input{glyphtounicode}\pdfgentounicode=1\fi   
\usepackage[T1]{fontenc}
\usepackage{lmodern}
\usepackage{booktabs}
\usepackage{graphicx}
\usepackage{url}
\usepackage[colorlinks=true,allcolors=blue]{hyperref}

\newcommand{\HtwoThrEps}{5.9\times10^{-3}}
\newcommand{\LiHThrEps}{5.5\times10^{-3}}
\newcommand{\GroverThrEps}{5.6\times10^{-3}}
\newcommand{\QPEThrEps}{9.0\times10^{-3}}
\newcommand{\ket}[1]{\lvert #1\rangle}
\newcommand{\Hz}{\mathrm{H}_2}
\newcommand{\tsw}{t_{\mathrm{SWAP}}}
\newcommand{\Sv}{\mathbf{S}}
\newcommand{\Popt}{P_{\mathrm{opt}}}

\begin{document}

\title{Pulse-timing errors in exchange-only spin qubits: how one control imperfection propagates to VQE, QAOA and fault-tolerance primitives}

\author{Yuichiro Minato}
\email{minato@blueqat.co.jp}
\affiliation{blueqat Inc., 2-24-12 Shibuya, Shibuya-ku, Tokyo 150-6139, Japan}

\date{\today}

\begin{abstract}
In an exchange-only (EO) spin qubit every gate is a sequence of Heisenberg
exchange pulses whose only continuous parameter is the pulse area. We ask how an error in that area propagates to an algorithm's result. On up to six encoded qubits we compile VQE for $\Hz$ and LiH, QAOA
MaxCut, Grover search, phase estimation and a distance-3 repetition code into
exchange pulses and perturb every pulse with a systematic fractional error
$\varepsilon$, an offset or per-pulse jitter. For circuits whose first-order response is negligible at the threshold, the useful regime ends at
$\varepsilon$ of a few $10^{-3}$: chemical accuracy is lost at
$\approx6\times10^{-3}$ for both molecules and $5\%$ of success probability between $3$ and $9\times10^{-3}$ for 498--1216 pulses. A systematic error
accumulates as the square of the depth and biases the answer; jitter accumulates linearly and, within statistics, does not bias phase estimation. Leakage from inter-block pulses is a
third of the CNOT error, survives variational re-optimisation and a leakage-controlled entangler, and makes a leakage-blind repetition-code round worse than its uncorrected data by a factor $38$--$93$ at $\varepsilon=10^{-2}$ in three of four compilations; in the fourth the uncorrected data already fail. The compilation moves the thresholds by a factor of two to eight. Measured over-rotation and exchange noise put devices at the edge of this regime, and with the measured $3.5\,\mu$s $T_2^*$ a quasi-static magnetic gradient limits thousand-pulse circuits unless $J/h$ reaches several hundred MHz with continuous areas, because a duration grid of $2$--$5\%$ of the SWAP time starts to cost chemical accuracy. As a bit budget, a common-mode area error needs $8$--$10$ significant bits and a $6$-bit significand for the durations suffices in the tested circuits; half-precision words and picosecond timers exceed this; older devices had $6$--$8$ bits of analog stability.
\end{abstract}

\maketitle

\section{Introduction}

Exchange-only (EO) qubits encode one logical qubit in the $S=\tfrac12$
subspace of three electron spins and manipulate it with nothing but the
Heisenberg exchange interaction between neighbouring
spins~\cite{bacon2000,divincenzo2000,kempe2001}. No magnetic-field gradient and
no microwave drive is needed: a gate is a train of voltage pulses on barrier
gates, each of which turns $J$ on for a while. The price is that the simplest
logical operation is a sequence of several pulses, a two-qubit gate a sequence
of about twenty, and the only thing the control electronics can set is the
exchange angle $\theta_k=\int J_k\,dt/\hbar$ of each pulse $k$. The literature
has concentrated on finding sequences that are exact and
short~\cite{fong2011,zeuch2016,setiawan2014,chadwick2024} and on characterising gate-level error and leakage~\cite{andrews2019,weinstein2023}; a cryogenic CMOS (cryo-CMOS) controller that sets exchange angles digitally has recently operated a device configurable for up to 18 EO qubits~\cite{hrl2026}. On the control side, van Dijk \emph{et al.}~\cite{vandijk2019} derived electrical specifications (digital-to-analogue converter (DAC) resolution, jitter, bandwidth) for a target gate fidelity of single-spin qubits, and Bhattacharjee \emph{et al.}~\cite{bhattacharjee2026} asked how a low DAC resolution degrades a quantum neural network. The HRL work also simulates quasi-static, axis-dependent fractional angle errors through its error-correction circuits. What has not been done, to our knowledge, is to follow one imprecision of the pulse area of an EO qubit through a set of complete algorithms with different figures of merit, and to state what resolution --- in the bits a control system actually has --- that implies.

This paper asks that question for a variational eigensolver (VQE) on $\Hz$ and
LiH, the quantum approximate optimisation algorithm (QAOA) for MaxCut, Grover
search, quantum phase estimation (QPE) and one round of a repetition code. Each
is compiled into exchange pulses, every pulse duration is perturbed with a
controlled error, and the quantity the algorithm is supposed to deliver is
recorded. We distinguish a systematic (calibration) error, the same for all
pulses in a run, from a random per-pulse jitter, and within each class a
fractional error from an additive offset. The simulation is exact on the
physical spins ($3n$ spins for $n$ encoded qubits), so leakage out of the code
space is tracked, and the readout of the repetition code is modelled
physically, so a leaked qubit is read the way a singlet--triplet detector
would read it.

Sections~\ref{sec:eo}--\ref{sec:model} describe the encoding, the compiler and
the error models. Sections~\ref{sec:gates}--\ref{sec:order} give the results:
gate level, accumulation with depth, each algorithm, and the conditions under
which the response is quadratic. Section~\ref{sec:variants} repeats the key
runs with a different single-qubit compilation and with two other entangling
sequences. Section~\ref{sec:deco} adds a quasi-static magnetic gradient and
asks how the choice of $J$ trades timing against dephasing.
Section~\ref{sec:hardware} confronts the thresholds with measured device
parameters, and Section~\ref{sec:limits} lists what the study does not
establish.

\section{Exchange-only qubits and their pulse sequences}
\label{sec:eo}

\subsection{Encoding, code space and leakage}

Encoded qubit $k$ occupies spins $(3k,3k+1,3k+2)$ of a linear chain. The
logical states are the two states of the three spins with total spin
$S=\tfrac12$ and projection $m=+\tfrac12$,
\begin{equation}
 \ket{0_L}=\ket{S}_{12}\otimes\ket{\uparrow},\qquad
 \ket{1_L}=\tfrac{1}{\sqrt6}\bigl(\ket{\uparrow\downarrow\uparrow}+\ket{\downarrow\uparrow\uparrow}
 -2\ket{\uparrow\uparrow\downarrow}\bigr),
\end{equation}
where $\ket S_{12}=(\ket{\uparrow\downarrow}-\ket{\downarrow\uparrow})/\sqrt2$ is the singlet of the first two spins; in $\ket{1_L}$ the
first two spins are in a triplet, so the pair $(1,2)$ is a singlet in
$\ket{0_L}$ and a triplet in $\ket{1_L}$, which makes a pulse on that pair
diagonal and its singlet--triplet readout a $Z$ measurement. Spins are numbered $0$ to $3n-1$, block $k$ holding $3k$, $3k{+}1$, $3k{+}2$; within a block, or a pair of blocks, we also write $1,2,3$ or $1,\dots,6$. The gauge of every block is taken to be polarised, a modelling assumption: the devices of Refs.~\cite{andrews2019,hrl2026} operate with an uninitialised, mixed gauge, and quantitative predictions for them would require averaging over gauge states. Exchange conserves
the total spin of the register and its projection $M$, so the computation stays in the
sector with $n$ spins down, of dimension $\binom{3n}{n}$ (18\,564 for $n=6$),
and we store the state in that sector only. In this sector $M=n/2$ is the
largest value compatible with every block having $S=\tfrac12$; a block cannot
change its gauge without another block acquiring $S=\tfrac32$. The code space
is therefore exactly the set of states in which every block has $S=\tfrac12$,
and ``leakage'' always means that at least one block has $S=\tfrac32$. Pulses
within a block conserve the block's spin and cannot leak; pulses between
blocks --- the entangling sequence and routing SWAPs --- can.

A pulse on spins $i,j$ of area $\theta=\pi t$ (durations $t$ in units of
$\tsw=\pi\hbar/J$) is
\begin{equation}
 U_{ij}(t)=e^{-i\pi t\,\Sv_i\cdot\Sv_j}
 \;\propto\;\cos\tfrac{\pi t}{2}\,\mathbf 1-i\sin\tfrac{\pi t}{2}\,\mathrm{SWAP}_{ij},
 \label{eq:pulse}
\end{equation}
i.e.\ $\mathrm{diag}(1,e^{-i\pi t})$ in the singlet/triplet basis of the pair,
the convention of Ref.~\cite{zeuch2016}; $t=1$ is a SWAP, $t=\tfrac12$ a
$\sqrt{\mathrm{SWAP}}$. Nominal durations are taken in $[0,2)$: a negative angle
is a long pulse, which matters once the error is proportional to the duration.

\subsection{Gates, routing, readout}

A pulse on the pair $(3k,3k{+}1)$ equals $R_z(-\pi t)$ up to a phase; a
pulse on $(3k{+}1,3k{+}2)$ rotates by the same angle about an axis $120^\circ$ from
$z$. $R_z$ therefore costs one pulse. Any other $SU(2)$ element is realised by
four pulses alternating between the two pairs, found by a least-squares search
from 16 random starts, keeping the shortest exact solution found (a heuristic;
Section~\ref{sec:variants} replaces it by fixed sequences).

For the entangling gate we implement the analytic construction of Zeuch and
Bonesteel~\cite{zeuch2016}, which reproduces the Fong--Wandzura
sequence~\cite{fong2011} up to single-qubit rotations. For the present
encoding it is the construction read with the order of the six spins reversed:
with the control on spins $4$--$6$ and the target on $1$--$3$, $R\,U_{45}(1)\,R\,U_{45}(1)\,R$ with
$R=U_{34}(\tfrac32)U_{23}(\tfrac12)U_{12}(1)U_{34}(1)U_{23}(\tfrac32)U_{34}(\tfrac12)$
(right to left): 20 pulses, eight SWAPs, six $\sqrt{\mathrm{SWAP}}$ and six
inverse $\sqrt{\mathrm{SWAP}}$. In the basis $\ket{c\,t}$ it acts on the code
space as $\mathrm{diag}(\mathbf 1,\hat n\cdot\boldsymbol\sigma)$ up to a global phase, with
$\hat n=(0,-\tfrac{\sqrt3}{2},-\tfrac12)$, with leakage below $10^{-15}$; $W$ is
the target rotation taking $\hat n$ to $\hat x$, so that
$(\mathbf 1\otimes W)\,G\,(\mathbf 1\otimes W^\dagger)$ is a CNOT (fidelity
$1.000$), 28 pulses in all (the exchange-pulse-optimizer toolchain~\cite{epo} also budgets 28 per CNOT). A CNOT whose control is
the left block is obtained from the same sequence by conjugating both qubits
with $H$, which exchanges control and target. The compiler merges all
single-qubit unitaries between two entangling gates on the same qubit into one
block, so the $W$ and $H$ corrections are usually absorbed. Section~\ref{sec:variants}
uses two further entangling sequences from the library of Chadwick \emph{et
al.}~\cite{chadwick2024,chadwickdata}.

Encoded qubits live on a line; a two-qubit gate between non-adjacent qubits is
served by moving one of them with logical SWAPs, each the physical exchange of
two blocks by nine adjacent $t=1$ pulses. For QAOA the moved qubit stays at its
new position; for Grover and QPE the compiler moves the target next to the
control and back. Pulse counts quoted below are those of the compiled circuits.

The physical readout of qubit $k$ is singlet-minus-triplet on its pair,
$Z_k^{\mathrm{phys}}=P_S-P_T=-\mathrm{SWAP}_{3k,3k+1}$, defined on leaked
states as well: a block with $S=\tfrac32$ has the pair in a triplet and reads
as $\ket1$. $X$ and $Y$ observables are obtained by adding the basis-change rotations to the circuit (the compiler merges them into the last single-qubit block of the qubit), with their own timing errors, before the $Z$ readout.
Expectation values are exact (no shot noise). The VQE energy is the sum of such
physically measured terms, each from its own compiled and perturbed circuit;
it is the number a VQE loop on this device would minimise, not the expectation
value of $H$ in one prepared state, and we call it the \emph{measured energy}.
For Grover, QPE and the QAOA optimal-cut probability we use the probabilities of the code-space amplitudes, which treat leaked population as failure and are what an experiment with leakage detection would report; the QAOA approximation ratio and the repetition code use the physical readout of every pair.

\section{Error models and figures of merit}
\label{sec:model}

Every pulse of nominal duration $t$ is executed with duration
\begin{equation}
 \begin{gathered}
 t'=t\,(1+\varepsilon)+\delta+t\,\sigma_{\mathrm m}\xi+\sigma_{\mathrm a}\xi',\\
 \xi,\xi'\sim\mathcal N(0,1)\text{ i.i.d.\ per pulse},
 \end{gathered}
\end{equation}
one term at a time. $\varepsilon$ and $\sigma_{\mathrm m}$ are dimensionless
fractional errors of the pulse area (a miscalibration of $J$ or of the clock,
or its per-pulse fluctuation); $\delta$ and $\sigma_{\mathrm a}$ are offsets in
units of $\tsw$ (a rise/fall-time error, or its fluctuation). Systematic errors
are scanned at $\pm10^{-4}\dots10^{-1}$; jitter at the same magnitudes,
averaged over independent draws of the whole sequence (100 for gates, $\Hz$ and the depth scan, 40 for the other circuits, 20 for the bond-dependent model). Sections~\ref{sec:deco}--\ref{sec:hardware} add a
duration grid, a bond-dependent static error and a magnetic gradient.

For each gate we compute the block $A$ of the physical operation projected on
the code space of dimension $d$, one error realisation per block (jitter
metrics averaged over 100 realisations), and report the coherent infidelity
$1-|\mathrm{tr}(U^\dagger A)|^2/d^2$ and the leakage $1-\mathrm{tr}(A^\dagger A)/d$.
For each algorithm we report a signed deviation from the ideal device, so
that an accidental improvement is visible: for VQE the measured energy minus
the exact ground energy $E_0$ of the qubit Hamiltonian (threshold
$1.6\,\mathrm{mHa}$, the usual chemical-accuracy figure applied to the
active-space energy); for QAOA the relative loss of the probability of an
optimal cut (threshold $5\%$), with the measured approximation ratio reported
alongside; for Grover and QPE the loss of success probability (threshold
$0.05$); for the repetition code the probability of a logical flip after one
round with physical readout and ideal decoding (threshold $10^{-3}$). The
threshold is the error at which the signed deviation first reaches the stated
value, interpolated between scanned levels.

\section{Gate level}
\label{sec:gates}

\begin{figure*}[t]
\centering
\includegraphics[width=\textwidth]{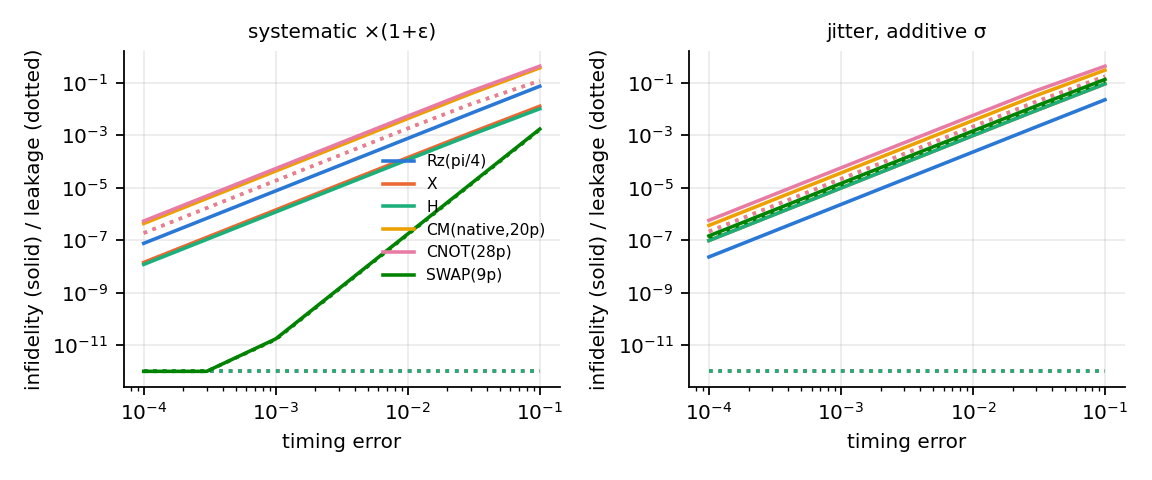}
\caption{Gate-level infidelity (solid) and leakage (dotted) versus timing
error for a systematic fractional error (left) and additive jitter (right).}
\label{fig:gates}
\end{figure*}

Table~\ref{tab:gates} gives the error of each primitive at an error of $10^{-2}$ in each model and
Fig.~\ref{fig:gates} the scans. Three observations follow.

\begin{table*}[t]
\centering\small
\caption{Primitive gates at a timing error of $10^{-2}$ ($\varepsilon$
fractional; $\delta$, $\sigma$ in units of $\tsw$): coherent infidelity $/$
leakage. $T$ is the total nominal duration in units of $\tsw$.}
\label{tab:gates}
\resizebox{\textwidth}{!}{\begin{tabular}{lrrcccccc}
\toprule
 & pulses & $T$ & \multicolumn{2}{c}{$\varepsilon=10^{-2}$} & \multicolumn{2}{c}{$\delta=10^{-2}$} & \multicolumn{2}{c}{$\sigma_{\mathrm a}=10^{-2}$}\\
 & & & infid. & leak & infid. & leak & infid. & leak\\
\midrule
$R_z(\pi/4)$ & 1 & 1.75 & $7.6\times10^{-4}$ & $1.0\times10^{-16}$ & $2.5\times10^{-4}$ & $1.0\times10^{-16}$ & $2.3\times10^{-4}$ & $1.0\times10^{-16}$ \\
$X$ & 4 & 2.04 & $1.4\times10^{-4}$ & $1.0\times10^{-16}$ & $4.0\times10^{-4}$ & $1.0\times10^{-16}$ & $9.7\times10^{-4}$ & $1.0\times10^{-16}$ \\
$H$ & 4 & 2.05 & $1.2\times10^{-4}$ & $1.0\times10^{-16}$ & $5.4\times10^{-4}$ & $2.2\times10^{-16}$ & $9.7\times10^{-4}$ & $1.0\times10^{-16}$ \\
native $C(\hat n\cdot\sigma)$ & 20 & 20.00 & $4.3\times10^{-3}$ & $1.8\times10^{-3}$ & $3.7\times10^{-3}$ & $1.3\times10^{-3}$ & $3.6\times10^{-3}$ & $2.2\times10^{-3}$ \\
CNOT & 28 & 24.54 & $5.3\times10^{-3}$ & $1.8\times10^{-3}$ & $4.3\times10^{-3}$ & $1.3\times10^{-3}$ & $5.7\times10^{-3}$ & $2.1\times10^{-3}$ \\
logical SWAP & 9 & 9.00 & $1.7\times10^{-7}$ & $1.6\times10^{-7}$ & $1.7\times10^{-7}$ & $1.6\times10^{-7}$ & $1.5\times10^{-3}$ & $1.3\times10^{-3}$ \\
\bottomrule
\end{tabular}}
\end{table*}

\emph{Single-qubit gates do not leak.} Exchange within a block conserves the
block's spin; a wrong duration rotates the qubit but cannot take it out of the code space. At $10^{-2}$ the four-pulse $X$ and $H$ have infidelity $1.2$--$1.4\times10^{-4}$, and the single $1.75$-$\tsw$ pulse of $R_z(\pi/4)$ $7.6\times10^{-4}$.

\emph{Pulses between blocks leak, and leakage is $30$--$42\%$ of their error under a systematic error (up to $60\%$ under jitter).} At $\varepsilon=10^{-2}$ the native entangling sequence has infidelity
$4.3\times10^{-3}$ of which $1.8\times10^{-3}$ is population outside the code
space; for the CNOT the fraction is one third. Under jitter the logical SWAP
leaks as well ($1.3\times10^{-3}$ at $\sigma_{\mathrm a}=10^{-2}$).

\emph{The logical SWAP is first-order immune to a common-mode systematic
error.} The nine-pulse block exchange has infidelity $1.7\times10^{-7}$ at
$\varepsilon=10^{-2}$, four orders of magnitude below a CNOT, with slope 4.
Write each erroneous SWAP as $U(1+\varepsilon)=U(1)\,e^{-i\pi\varepsilon\,\Sv_i\cdot\Sv_j}$
and move all error factors through the remaining permutations: each
transposition crosses one spin $a$ of block $A$ with one spin $b$ of block $B$,
each pair crosses once, so to first order the error is
$1-i\pi\varepsilon\sum_{a\in A,b\in B}\Sv_a\cdot\Sv_b=1-i\pi\varepsilon\,\Sv_A\cdot\Sv_B$.
For two unleaked blocks with polarised gauge the pair has total spin $1$ and
$\Sv_A\cdot\Sv_B=\tfrac14$: a global phase. The argument needs the same
$\varepsilon$ on every bond; with independent static errors on the five bonds
(standard deviation $10^{-2}$) the SWAP's infidelity is $2.1\times10^{-3}$,
one fifth of a CNOT's under the same model, and it does not protect blocks that
have already leaked. Jitter breaks the cancellation as well. Routing unleaked
qubits is thus nearly free against a common calibration error and is paid for
in bond-to-bond differences and in jitter.

\section{Accumulation with depth}
\label{sec:depth}

\begin{figure*}[t]
\centering
\includegraphics[width=\textwidth]{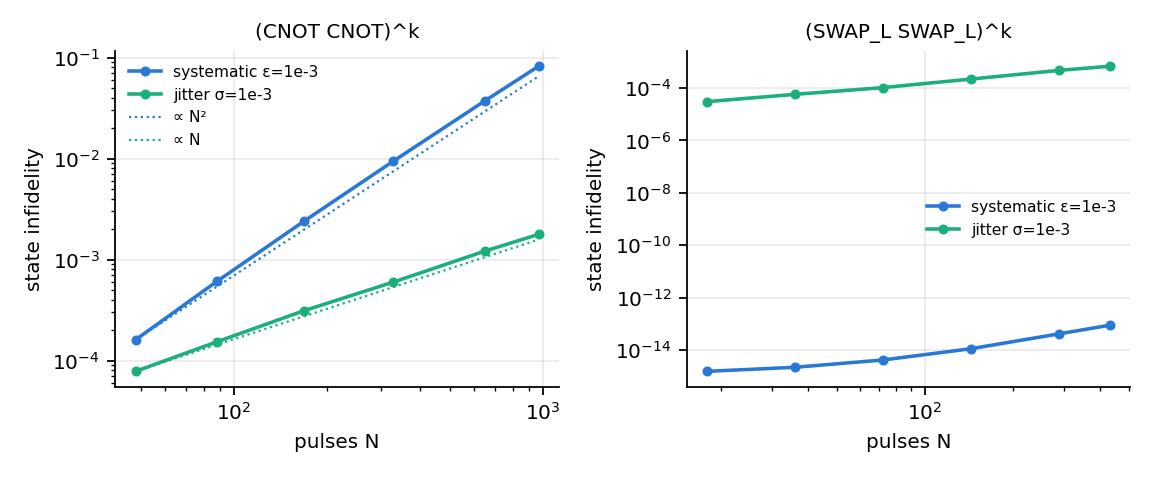}
\caption{State infidelity versus pulse count for a repeated identity.}
\label{fig:depth}
\end{figure*}

To separate the two error classes we repeat an identity, $(\mathrm{CNOT}\,
\mathrm{CNOT})^k$ and $(\mathrm{SWAP}_L\mathrm{SWAP}_L)^k$, with $k$ up to 24 at
$\varepsilon=\sigma_{\mathrm a}=10^{-3}$ (Table~\ref{tab:depth},
Fig.~\ref{fig:depth}). The systematic error grows as $N^2$ in the pulse count
$N$ and the jitter as $N$: over the 20-fold range of $N$ (24-fold in the number of entanglers, $24^2=576$) the former grows by a factor $512$ and the latter by $23$; at $10^{-3}$ and a thousand pulses the
same error is $46$ times more damaging when it is systematic. The SWAP chain
stays below $10^{-13}$ under the common-mode systematic error.

\begin{table*}[t]
\centering\small
\caption{Accumulation with depth at $\varepsilon=\sigma_{\mathrm a}=10^{-3}$:
state infidelity of $(\mathrm{CNOT}\,\mathrm{CNOT})^k$ and
$(\mathrm{SWAP}_L\mathrm{SWAP}_L)^k$ with the control in $\ket+$ and the target in $\ket0$; jitter over 100 draws.}
\label{tab:depth}
\begin{tabular}{rrccrcc}
\toprule
 & \multicolumn{3}{c}{$(\mathrm{CNOT}\,\mathrm{CNOT})^k$} & \multicolumn{3}{c}{$(\mathrm{SWAP}_L\mathrm{SWAP}_L)^k$}\\
$k$ & pulses & systematic & jitter & pulses & systematic & jitter\\
\midrule
1 & 48 & $1.6\times10^{-4}$ & $7.8\times10^{-5}$ & 18 & $1.6\times10^{-15}$ & $3.0\times10^{-5}$ \\
2 & 88 & $6.1\times10^{-4}$ & $1.5\times10^{-4}$ & 36 & $2.2\times10^{-15}$ & $5.8\times10^{-5}$ \\
4 & 168 & $2.4\times10^{-3}$ & $3.1\times10^{-4}$ & 72 & $4.2\times10^{-15}$ & $1.0\times10^{-4}$ \\
8 & 328 & $9.4\times10^{-3}$ & $6.0\times10^{-4}$ & 144 & $1.1\times10^{-14}$ & $2.2\times10^{-4}$ \\
16 & 648 & $3.7\times10^{-2}$ & $1.2\times10^{-3}$ & 288 & $4.2\times10^{-14}$ & $4.7\times10^{-4}$ \\
24 & 968 & $8.2\times10^{-2}$ & $1.8\times10^{-3}$ & 432 & $9.0\times10^{-14}$ & $6.8\times10^{-4}$ \\
\bottomrule
\end{tabular}
\end{table*}

This $N^2$ law describes repeating the same erroneous operation. It does not
carry over to a comparison between algorithms: the even coefficients of
Table~\ref{tab:linquad} grow far more slowly than $N^2$ across circuits (from $47$ for 32 pulses to $1.3\times10^3$ for 1216, a factor 28 for a factor 38 in $N$), because different gates rotate different things and their errors partly
cancel, and because the figures of merit differ in sensitivity. This is why,
for circuits without a first-order term and a given error model, the
thresholds of Table~\ref{tab:thr} vary by less than a factor of three across a nearly 40-fold
range of pulse count.

\section{Algorithms}
\label{sec:algos}

\subsection{VQE for $\Hz$ and LiH}

For $\Hz$ we use the two-qubit STO-3G Hamiltonian at $0.735$\,\AA{}
($II$, $IZ$, $ZI$, $ZZ$, $XX$; exact electronic ground energy $-1.85727503$\,Ha),
in the frame used by the Qiskit tutorials~\cite{qiskitH2}; we recomputed it
with PySCF through OpenFermion~\cite{openfermion} (symmetry-conserving
Bravyi--Kitaev reduction of the four-spin-orbital Hamiltonian) and obtained the same coefficients up to a flip of one qubit's $Z$ axis and the nuclear repulsion of $0.71997$\,Ha. The one-parameter ansatz $R_y(\theta)\,\mathrm{CNOT}\,X$ reaches the exact ground state: 32 pulses plus the basis change of the $XX$ term. For LiH we compute the integrals with PySCF through
OpenFermion~\cite{openfermion}, freeze the Li $1s$ orbital, keep the next three
spatial orbitals and two electrons, and reduce to four qubits with the
symmetry-conserving Bravyi--Kitaev mapping (52 Pauli terms including the identity, in 13 settings;
exact energy $-7.86291934$\,Ha at $1.6$\,\AA). A hardware-efficient ansatz
($R_y$ on every qubit, CNOTs $3{\to}2$, $2{\to}1$, $1{\to}0$, twice, then a
final $R_y$ layer; 12 parameters, 184 pulses) reaches that energy to
$10^{-11}$\,Ha. Parameters are optimised on the ideal device; for $\Hz$ the
parameter is re-optimised on the erroneous device at every systematic level,
for LiH at $\varepsilon,\delta=3\times10^{-3},10^{-2},3\times10^{-2}$ (positive
sign) with a local Nelder--Mead search.

\begin{figure*}[tp]
\centering
\includegraphics[width=\textwidth]{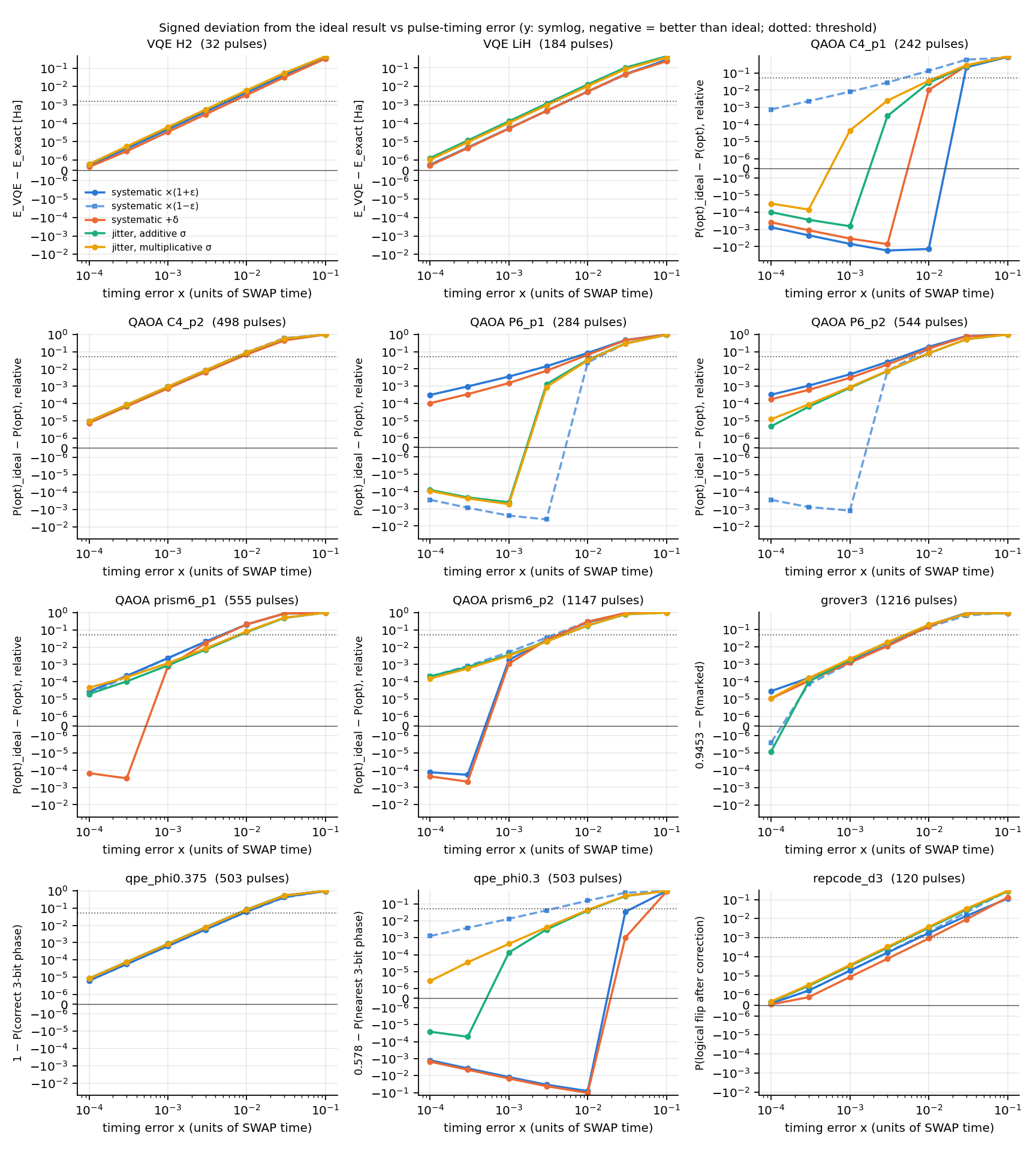}
\caption{Signed deviation from the ideal result versus timing error. Blue:
systematic fractional error, solid $+\varepsilon$, dashed $-\varepsilon$; red:
offset $+\delta$; green and orange: additive and multiplicative jitter. The
vertical axis is symmetric log; a result better than ideal appears below zero.
Dotted: the threshold of Table~\ref{tab:thr}. Where the $\pm\varepsilon$
curves coincide the response is even; where they split a first-order term is
present (Section~\ref{sec:order}). The smallest-$\sigma$ jitter points carry the Monte Carlo standard error of $40$--$100$ draws, which is first order in $\sigma$.}
\label{fig:main}
\end{figure*}

Both molecules behave alike (Fig.~\ref{fig:main}, Table~\ref{tab:thr}). The
measured energy error grows as the square of the error with no sign
dependence and crosses chemical accuracy at $\varepsilon=\HtwoThrEps$ for
$\Hz$ and $\LiHThrEps$ for LiH, although LiH has nearly six times the pulses;
we have not identified the mechanism, and note only that jitter, which cannot
cancel between gates, is more damaging for LiH ($\sigma_{\mathrm a}$ threshold
$3.5\times10^{-3}$ against $5.2\times10^{-3}$). Projecting the prepared state
on the code space, renormalising and measuring it ideally gives about $0.6$ of
the measured error at every level (Table~\ref{tab:decomp}); the remaining
$0.4$ is the price of leakage read as $\ket1$ and of imperfect basis changes.

\begin{table}[t]
\centering\small
\caption{Decomposition of the measured VQE energy error (Ha) at a systematic
fractional error $\varepsilon$: total, the part that remains when the prepared
state is projected on the code space and measured ideally, and the leaked
population.}
\label{tab:decomp}
\begin{tabular}{lcccc}
\toprule
 & $\varepsilon$ & $\Delta E$ measured & $\Delta E$ code space & leakage\\
\midrule
$\Hz$ & $10^{-3}$ & $4.7\times10^{-5}$ & $2.9\times10^{-5}$ & $2.3\times10^{-5}$\\
      & $10^{-2}$ & $4.7\times10^{-3}$ & $3.0\times10^{-3}$ & $2.3\times10^{-3}$\\
LiH   & $10^{-3}$ & $5.3\times10^{-5}$ & $3.4\times10^{-5}$ & $8.1\times10^{-5}$\\
      & $10^{-2}$ & $5.3\times10^{-3}$ & $3.4\times10^{-3}$ & $8.0\times10^{-3}$\\
\bottomrule
\end{tabular}
\end{table}

Re-optimisation recovers part of the coherent error and little of the
leakage. For $\Hz$ the parameter moves by $+0.002$, $+0.025$ and $+0.23$\,rad
at $\varepsilon=10^{-3}$, $10^{-2}$ and $3\times10^{-2}$; $\Delta E$ at
$10^{-2}$ falls from $4.7$ to $4.4$\,mHa (from $3.4$ to $2.7$\,mHa for
$\delta$), and the leaked population is unchanged to two significant digits
($2.264\times10^{-3}\to2.267\times10^{-3}$). For LiH the twelve parameters
halve the error, $5.3$ to $2.8$\,mHa, moving the re-optimised threshold to
roughly $8\times10^{-3}$; the leakage at the re-optimised LiH parameters was
not recorded. In QAOA, where it was recorded for $C_4$ and the prism graph,
re-optimisation changes the leakage by at most $7\times10^{-3}$ in absolute
terms up to $\varepsilon=10^{-2}$.

\subsection{QAOA MaxCut}
\label{sec:qaoa}

We run QAOA at $p=1$ and $2$ on the 4-cycle $C_4$ (for which $p=2$ reaches the
exact MaxCut state, $\Popt=1$), the 6-path $P_6$ (no routing) and the 6-vertex
prism graph (nine edges, routing on every layer); Table~\ref{tab:qaoa} gives the results.

\begin{table*}[t]
\centering\small
\caption{QAOA MaxCut: measured approximation ratio $/$ probability of an
optimal cut (code-space amplitudes) at a systematic fractional error
$\varepsilon$ (re-optimised ratio in parentheses) and at additive jitter
$\sigma_{\mathrm a}=10^{-2}$ (40 draws).}
\label{tab:qaoa}
\setlength{\tabcolsep}{4pt}
\begin{tabular}{lrccccc}
\toprule
graph, $p$ & pulses & ideal & $\varepsilon=3\times10^{-3}$ & $\varepsilon=10^{-2}$ & $\varepsilon=3\times10^{-2}$ & $\sigma_{\mathrm a}=10^{-2}$\\
\midrule
$C_4$, 1 & 242 & 0.750 / 0.53 & 0.754 / 0.54 & 0.751 / 0.54 (0.756) & 0.681 / 0.42 (0.720) & 0.744 / 0.52\\
$C_4$, 2 & 498 & 1.000 / 1.00 & 0.995 / 0.99 & 0.952 / 0.91 (0.966) & 0.720 / 0.49 (0.791) & 0.953 / 0.91\\
$P_6$, 1 & 284 & 0.774 / 0.24 & 0.772 / 0.24 & 0.761 / 0.22 (0.762) & 0.661 / 0.13 (0.683) & 0.767 / 0.23\\
$P_6$, 2 & 544 & 0.868 / 0.44 & 0.864 / 0.43 & 0.824 / 0.35 (0.840) & 0.583 / 0.08 (0.694) & 0.850 / 0.40\\
prism, 1 & 555 & 0.848 / 0.40 & 0.846 / 0.39 & 0.818 / 0.32 (0.824) & 0.669 / 0.05 (0.723) & 0.836 / 0.37\\
prism, 2 & 1147 & 0.943 / 0.72 & 0.936 / 0.70 & 0.869 / 0.52 (0.877) & 0.631 / 0.07 (0.739) & 0.904 / 0.59\\
\bottomrule
\end{tabular}
\end{table*}

Shallow QAOA is the most forgiving circuit here: at $p=1$ the ratio loses at
most $0.030$ at $\varepsilon=10^{-2}$, because it is an average over a broad
distribution and the circuit is short. Buying a sharper answer with $p=2$
costs that robustness: on $C_4$ the exact solution falls to $\Popt=0.91$ at
$10^{-2}$ and $0.49$ at $3\times10^{-2}$, and the 1147-pulse prism circuit
loses $28\%$ of its $\Popt$ at $10^{-2}$. The $5\%$ thresholds in $\Popt$ lie
between $3\times10^{-3}$ and $9\times10^{-3}$ at $p=2$ for all error models; at
$p=1$ they depend on the sign and model ($4.7\times10^{-3}$ to
$1.7\times10^{-2}$) because of the first-order term of
Section~\ref{sec:order}. Re-optimisation of $(\gamma,\beta)$ gains at most
$0.016$ in ratio up to $10^{-2}$ and $0.02$--$0.11$ at $3\times10^{-2}$.

Two remarks on metrics. On $C_4$ at $p=1$ the measured ratio \emph{exceeds}
the ideal $0.750$ for $0<\varepsilon\le10^{-2}$, and so does the ratio
conditioned on the code space ($0.7537$ against $0.7536$): a coherent
first-order effect of the over-rotation, not a readout artefact. And $\Popt$
as defined here rejects leaked outcomes; a plain singlet--triplet readout
counts some leaked outcomes as optimal cuts, so an experimental $\Popt$ needs
leakage detection or must be read as an upper bound.

\subsection{Grover search and phase estimation}

Grover: three qubits, marked state $\ket{101}$, two iterations (ideal success
$0.9453$), each CCZ as exponentials of $Z$ strings with one routed term, 1216
pulses. The $5\%$ threshold is $\GroverThrEps$; at $\varepsilon=10^{-2}$ the
success is $0.77$ with leakage $0.11$, at $3\times10^{-2}$ it is $0.05$ with
leakage $0.62$.

QPE: three counting qubits on a phase gate, target in $\ket1$, 503 pulses with
routed controlled phases. For $\phi=3/8$ the success falls from $1$ to $0.94$
at $10^{-2}$ and $0.57$ at $3\times10^{-2}$; threshold $\QPEThrEps$. For
$\phi=0.3$ the ideal device returns the nearest value $010$ with probability
$0.578$; for $\varepsilon>0$ the distribution of this compiled circuit slides towards $0.25$ and that probability rises (to $0.66$ at $10^{-2}$), while for $\varepsilon<0$ it falls (to $0.43$); every pulse of the preparation, routing and inverse transform is scaled, so the shift is a property of the compiled circuit rather than of the controlled phases alone. The circular mean of the estimated phase (Table~\ref{tab:qpe})
shifts linearly, $\Delta\phi\simeq-2\varepsilon$, while the distribution stays
sharp. Under additive jitter of $10^{-2}$ the mean is $0.286\pm0.002$ (40
draws) against $0.285$ ideal, unchanged within statistics, while a systematic
error of the same size moves it to $0.267$ or $0.308$. A systematic timing
error changes the answer of QPE; jitter blurs it.

\begin{table*}[t]
\centering\small
\caption{QPE with $\phi=0.3$: circular mean of the estimated phase versus
$\varepsilon$ (ideal device $0.285$).}
\label{tab:qpe}
\begin{tabular}{lccccccccc}
\toprule
$\varepsilon$ & $-0.03$ & $-0.01$ & $-3.0\times10^{-3}$ & $-1.0\times10^{-3}$ & $0$ & $+1.0\times10^{-3}$ & $+3.0\times10^{-3}$ & $+0.01$ & $+0.03$ \\
$\bar\phi$ & 0.369 & 0.308 & 0.291 & 0.287 & 0.285 & 0.283 & 0.279 & 0.267 & 0.241 \\
\bottomrule
\end{tabular}
\end{table*}

\subsection{Repetition code}

Three data qubits in $\ket{000}$ interleaved with two ancillas, one round of
four CNOTs (120 pulses), physical readout of all five pairs, ideal decoding of
the ancilla readout and majority vote of the corrected data. The logical flip
after correction is $1.6\times10^{-4}$ at $\varepsilon=3\times10^{-3}$,
$1.7\times10^{-3}$ at $10^{-2}$ and $1.4\times10^{-2}$ at $3\times10^{-2}$:
second order, like a single gate. The same noisy syndrome-extraction circuit, with its data read \emph{without} applying the correction, gives a majority flip of only $1.9\times10^{-5}$ at $10^{-2}$. Applying the correction makes things worse by two orders of magnitude. A single false
syndrome cannot defeat a majority vote, so the mechanism must be correlated,
and it is: $99\%$ of the failures at $10^{-2}$ are the single readout pattern in which the middle data qubit reads $1$ and the syndrome is $10$; $94\%$ of that pattern's weight lies outside the code space; the CNOT between the middle data qubit and its ancilla leaks both blocks together (joint probability of both reading as triplets $3.1\times10^{-3}$, against $2.4\times10^{-5}$ if independent; see \texttt{extra\_checks.json} in the repository). The leaked data block reads as $1$, the leaked ancilla announces
a syndrome, and the decoder flips the neighbouring data qubit that was
correct. With leaked outcomes discarded before decoding the corrected flip is
$1\times10^{-4}$. For EO qubits with timing error, leakage, not the logical flip, sets the budget of this code with this decoder, and a decoder that cannot see leakage increased the logical error in our simulation; the result motivates leakage-aware decoding or leakage-reduction units~\cite{fong2011}, which we did not simulate.

\section{When is the response quadratic?}
\label{sec:order}

Table~\ref{tab:linquad} separates the odd part $b=[e(+x)-e(-x)]/2x$ from the
even part $a=[e(+x)+e(-x)]/2x^2$ of the response at $x=10^{-3}$ and gives the
crossover $x_c=|b|/a$ below which the odd part dominates. These are
finite-difference estimates at one level, not Taylor coefficients (for the
quartic logical SWAP $a$ is not a second derivative at all); the ideal device
has zero error for every figure of merit, so no baseline is subtracted.

\begin{table}[t]
\centering\small
\caption{Odd and even parts of the response to a systematic fractional error
from the pair $\varepsilon=\pm10^{-3}$. ``Ideal'' is the figure of merit on
the ideal device.}
\label{tab:linquad}
\setlength{\tabcolsep}{3pt}
\resizebox{\columnwidth}{!}{\begin{tabular}{lcccc}
\toprule
 & ideal & $b$ & $a$ & $x_c=|b|/a$\\
\midrule
VQE H$_2$ (2q) & $E_0$ & $-3.3\times10^{-6}$ & $4.7\times10^{1}$ & $<10^{-7}$ \\
VQE LiH (4q) & $E_0$ & $+4.3\times10^{-4}$ & $5.3\times10^{1}$ & $8.2\times10^{-6}$ \\
QAOA $C_4$, $p=1$ & 0.531 & $-7.5$ & $6.3\times10^{2}$ & $1.2\times10^{-2}$ \\
QAOA $C_4$, $p=2$ & 1.000 & $-4.7\times10^{-3}$ & $9.4\times10^{2}$ & $5.0\times10^{-6}$ \\
QAOA $P_6$, $p=1$ & 0.239 & $+3$ & $5.6\times10^{2}$ & $5.4\times10^{-3}$ \\
QAOA $P_6$, $p=2$ & 0.437 & $+3.1$ & $1.9\times10^{3}$ & $1.6\times10^{-3}$ \\
QAOA prism, $p=1$ & 0.401 & $+0.035$ & $2.3\times10^{3}$ & $1.5\times10^{-5}$ \\
QAOA prism, $p=2$ & 0.721 & $-1.7$ & $3.5\times10^{3}$ & $4.8\times10^{-4}$ \\
Grover (3q, 2 iter.) & 0.9453 & $+0.14$ & $1.3\times10^{3}$ & $1.1\times10^{-4}$ \\
QPE, $\phi=3/8$ & 1 & $+8.9\times10^{-4}$ & $6.3\times10^{2}$ & $1.4\times10^{-6}$ \\
QPE, $\phi=0.3$ & 0.5775 & $-13$ & $3.7\times10^{2}$ & $3.4\times10^{-2}$ \\
Repetition code $d=3$ & 0 & $-1.4\times10^{-4}$ & $1.8\times10^{1}$ & $7.4\times10^{-6}$ \\
\bottomrule
\end{tabular}}
\end{table}

Wherever the ideal value is an extremum of the figure of merit --- the exact
ground energy, $\Popt=1$ for $C_4$ at $p=2$, success $1$ for exact QPE, flip
probability $0$ for the code, fidelity $1$ for every gate --- a first-order term is excluded (for the measured VQE energy also because leakage is second order: a leaked block can read below $E_0$); the leading order is then even but not necessarily second
(the SWAP is quartic), the odd part is at the precision of the finite
difference ($x_c\lesssim10^{-5}$), and the $\pm\varepsilon$ curves in
Fig.~\ref{fig:main} coincide. Wherever it is not, a first-order term is
allowed and usually present: for $x<x_c$ ($5\times10^{-3}$ for $P_6$ and
$1.2\times10^{-2}$ for $C_4$ at $p=1$, $3\times10^{-2}$ for QPE at $\phi=0.3$) an
error of one sign improves the result. Grover's $0.945$ is not an extremum
but its odd part is small ($x_c=1.1\times10^{-4}$ with the present
compilation, $10^{-3}$ with an earlier one); $P_6$ and the prism graph at $p=2$
have $x_c=1.6\times10^{-3}$ and $4.8\times10^{-4}$, well below their
thresholds; the prism graph at $p=1$ has $x_c=1.5\times10^{-5}$. The absence
of an extremum permits a first-order term; it does not force one, and its
size depends on the compiled circuit. For VQE the argument applies to the
prepared state; the measured energy also contains the perturbed basis
changes, which could add an odd term, but none is seen. Jitter has zero mean,
so its odd part averages out over the ensemble (the shot-to-shot spread, not
analysed here, is first order in $\sigma$); with 40--100 draws a residual of first order in $\sigma$ remains (the Monte Carlo standard error), which is why the smallest-$\sigma$ points of the first-order circuits scatter around zero.

\begin{table*}[t]
\centering\small
\caption{Timing error at which each figure of merit reaches its threshold
(chemical accuracy for VQE, $5\%$ relative loss of $\Popt$ for QAOA, $0.05$
loss of success for Grover and QPE, $10^{-3}$ logical flip for the code).
$\varepsilon$, $\sigma_{\mathrm m}$ fractional; $\delta$, $\sigma_{\mathrm a}$
in units of $\tsw$. Entries for circuits with a first-order term ($C_4$ and
$P_6$ at $p=1$, QPE at $\phi=0.3$) depend on the sign.}
\label{tab:thr}
\begin{tabular}{lrccccc}
\toprule
 & pulses & $+\varepsilon$ & $-\varepsilon$ & $+\delta$ & $\sigma_{\mathrm a}$ & $\sigma_{\mathrm m}$\\
\midrule
VQE H$_2$ (2q) & 32 & $5.9\times10^{-3}$ & $5.9\times10^{-3}$ & $6.8\times10^{-3}$ & $5.2\times10^{-3}$ & $5.1\times10^{-3}$ \\
VQE LiH (4q) & 184 & $5.5\times10^{-3}$ & $5.6\times10^{-3}$ & $5.5\times10^{-3}$ & $3.5\times10^{-3}$ & $3.9\times10^{-3}$ \\
QAOA $C_4$, $p=1$ & 242 & $1.6\times10^{-2}$ & $4.7\times10^{-3}$ & $1.7\times10^{-2}$ & $1.4\times10^{-2}$ & $1.2\times10^{-2}$ \\
QAOA $C_4$, $p=2$ & 498 & $7.6\times10^{-3}$ & $7.3\times10^{-3}$ & $8.5\times10^{-3}$ & $7.5\times10^{-3}$ & $7.4\times10^{-3}$ \\
QAOA $P_6$, $p=1$ & 284 & $7.0\times10^{-3}$ & $1.4\times10^{-2}$ & $8.6\times10^{-3}$ & $1.2\times10^{-2}$ & $1.3\times10^{-2}$ \\
QAOA $P_6$, $p=2$ & 544 & $4.5\times10^{-3}$ & $6.4\times10^{-3}$ & $5.3\times10^{-3}$ & $7.8\times10^{-3}$ & $7.7\times10^{-3}$ \\
QAOA prism, $p=1$ & 555 & $4.8\times10^{-3}$ & $4.7\times10^{-3}$ & $5.0\times10^{-3}$ & $8.2\times10^{-3}$ & $7.7\times10^{-3}$ \\
QAOA prism, $p=2$ & 1147 & $4.2\times10^{-3}$ & $3.6\times10^{-3}$ & $4.1\times10^{-3}$ & $4.9\times10^{-3}$ & $4.9\times10^{-3}$ \\
Grover (3q, 2 iter.) & 1216 & $5.6\times10^{-3}$ & $6.1\times10^{-3}$ & $6.0\times10^{-3}$ & $5.1\times10^{-3}$ & $5.0\times10^{-3}$ \\
QPE, $\phi=3/8$ & 503 & $9.0\times10^{-3}$ & $9.1\times10^{-3}$ & $7.8\times10^{-3}$ & $7.8\times10^{-3}$ & $7.9\times10^{-3}$ \\
QPE, $\phi=0.3$ & 503 & $3.6\times10^{-2}$ & $3.6\times10^{-3}$ & $6.4\times10^{-2}$ & $1.1\times10^{-2}$ & $1.1\times10^{-2}$ \\
Repetition code $d=3$ & 120 & $7.6\times10^{-3}$ & $7.1\times10^{-3}$ & $1.0\times10^{-2}$ & $5.5\times10^{-3}$ & $5.2\times10^{-3}$ \\
\bottomrule
\end{tabular}
\end{table*}

\section{Dependence on the compilation}
\label{sec:variants}

Two choices above are ours rather than the hardware's: the four-pulse
single-qubit decompositions found by a numerical search, re-chosen whenever a
merged rotation changes (two circuits equal on an ideal device can compile to
sequences whose errors differ; we saw a $17\%$ difference in the $\Hz$ energy error between
two search settings), and the Zeuch--Bonesteel entangler. We therefore
repeated the key runs with three alternatives. (i) \emph{Fixed} single-qubit sequences: $R_z$ is one pulse and every other rotation, written as $R_z(\alpha)R_y(\beta)R_z(\gamma)$, is realised as $R_z(\alpha)\,[S\,H\,R_z(\beta)\,H\,S^\dagger]\,R_z(\gamma)$ with one fixed four-pulse block for $H$, so that a rotation costs 11--13 pulses (12 for $H$ itself) and a CNOT 46. (ii) The 26-pulse CZ and (iii) the 44-pulse leakage-controlled CZ (LCCZ)
of the linear six-dot entry of the Chadwick \emph{et al.}
library~\cite{chadwickdata}. Their angle convention $U=\cos\tfrac\theta2+
i\sin\tfrac\theta2\,\mathrm{SWAP}$ maps to ours as $t=2-\theta/\pi$, and their
qubit basis (singlet on the first pair of each block) is ours, so both sequences
are used as published and are a CZ in our basis (exact for the 26-pulse sequence, infidelity $4\times10^{-8}$ for the leakage-controlled one; leakage below $10^{-14}$); $\mathrm{CNOT}=H_t\,\mathrm{CZ}\,H_t$ for target $t$, 34 and 52
pulses before merging. Table~\ref{tab:var}
compares.

\begin{table*}[t]
\centering\small
\caption{Robustness to the compilation. Top: CNOT at a timing error of
$10^{-2}$ for the main compilation, fixed single-qubit sequences, and the two
Chadwick entanglers. Bottom: pulse count, $+\varepsilon$ threshold and
leakage at $\varepsilon=10^{-2}$ per circuit, threshold and leakage in units of $10^{-3}$. LiH and QAOA were not repeated with
the Chadwick entanglers.}
\label{tab:var}
\begin{tabular}{lrrcccccc}
\toprule
CNOT & pulses & $T$ & infid.\ $\varepsilon$ & leak & fraction & infid.\ $\sigma_{\mathrm a}$ & leak\\
\midrule
main & 28 & 24.5 & $5.3\times10^{-3}$ & $1.8\times10^{-3}$ & 34\% & $5.7\times10^{-3}$ & $2.1\times10^{-3}$ \\
fixed 1q & 46 & 37.7 & $1.5\times10^{-2}$ & $1.8\times10^{-3}$ & 12\% & $9.8\times10^{-3}$ & $2.2\times10^{-3}$ \\
CZ (Chadwick) & 34 & 34.1 & $5.1\times10^{-3}$ & $2.5\times10^{-3}$ & 48\% & $7.1\times10^{-3}$ & $2.2\times10^{-3}$ \\
LCCZ (Chadwick) & 52 & 49.6 & $1.2\times10^{-2}$ & $4.1\times10^{-3}$ & 35\% & $1.1\times10^{-2}$ & $4.0\times10^{-3}$ \\
\bottomrule
\end{tabular}

\medskip
{\setlength{\tabcolsep}{3.5pt}
\begin{tabular}{lccc|ccc|ccc|ccc}
\toprule
 & \multicolumn{3}{c|}{main} & \multicolumn{3}{c|}{fixed 1q} & \multicolumn{3}{c|}{CZ} & \multicolumn{3}{c}{LCCZ}\\
 & $N$ & thr. & leak & $N$ & thr. & leak & $N$ & thr. & leak & $N$ & thr. & leak\\
\midrule
VQE H$_2$ & 32 & 5.9 & 2.3 & 59 & 3.1 & 2.3 & 38 & 6.2 & 2.6 & 56 & 3.9 & 5.6 \\
VQE LiH & 184 & 5.5 & 8.0 & 328 & 3.5 & 6.5 & --- & --- & --- & --- & --- & --- \\
QAOA $C_4$ $p{=}2$ & 498 & 7.6 & 44 & 776 & 4.4 & 49 & --- & --- & --- & --- & --- & --- \\
QAOA prism $p{=}2$ & 1147 & 4.2 & 89 & 1721 & 2.5 & 102 & --- & --- & --- & --- & --- & --- \\
Grover & 1216 & 5.6 & 113 & 1817 & 2.9 & 181 & 1450 & 4.9 & 71 & 2170 & 3.3 & 226 \\
QPE $3/8$ & 503 & 9.0 & 26 & 746 & 5.1 & 20 & 561 & 9.0 & 32 & 777 & 8.6 & 43 \\
Rep.\ code & 120 & 7.6 & 6.7 & 208 & 5.0 & 7.0 & 120 & 6.0 & 14 & 192 & 4.2 & 20 \\
\bottomrule
\end{tabular}}
\end{table*}

The qualitative picture --- thresholds of a few $10^{-3}$, leakage from inter-block pulses that nothing downstream repairs --- does not move, but the numbers do. With fixed single-qubit sequences every circuit has $1.5$--$1.8$ times the pulses and the $+\varepsilon$ thresholds fall by a factor $1.5$--$2$ ($1.3$--$2.3$ over all circuits; $\Hz$ $5.9\to3.1\times10^{-3}$, LiH $5.5\to3.5\times10^{-3}$, Grover $5.6\to2.9\times10^{-3}$). The CNOT's own leakage is unchanged, because the extra pulses are all within blocks, so its leakage fraction drops from a third to $12\%$; at the circuit level the leakage at $\varepsilon=10^{-2}$ moves in either direction (LiH $8.0\to6.5\times10^{-3}$, Grover $0.11\to0.18$, QPE $0.026\to0.020$, and up to a factor five for the shallow $p=1$ QAOA circuits, which are not in the table and whose leakage is small), because the extra rotations change the states that enter the entanglers. With the Chadwick
CZ the CNOT has $34$ pulses, infidelity $5.1\times10^{-3}$ and leakage
$2.5\times10^{-3}$ ($48\%$); Grover's threshold falls to $4.9\times10^{-3}$
while QPE's is unchanged ($9.0\times10^{-3}$). The leakage-controlled CZ, designed to stop leakage from \emph{spreading} between qubits, does not reduce the leakage \emph{produced} by a timing error: $4.1\times10^{-3}$ ($35\%$ of $1.2\times10^{-2}$) at the gate level and $0.23$ for Grover at $\varepsilon=10^{-2}$ against $0.11$ with the 28-pulse CNOT. The repetition code is where the entangler matters most. The ratio of corrected to uncorrected logical flip at $\varepsilon=10^{-2}$ is $93$ (main), $38$ (fixed sequences) and $54$ (leakage-controlled CZ), but only $1.2$ with the Chadwick CZ: there the corrected flip, $2.8\times10^{-3}$, is no better than the main compilation's, while the uncorrected data already fail at $2.3\times10^{-3}$ (against $1.9\times10^{-5}$ with the Zeuch--Bonesteel sequence), because the symmetric CZ sequence leaks both blocks of a CNOT and the two data blocks that share an ancilla then read as $1$ together (the pattern $d=110$, $s=10$ carries $1.0\times10^{-3}$ and lies almost entirely outside the code space). At $\varepsilon\le10^{-2}$ leakage-blind correction helps in none of the four compilations; how much it hurts depends on which blocks the sequence leaks. The three sequences have $9$, $6$ and $10$ inter-block pulses, so the leakage at a given $\varepsilon$ is a property of the sequence rather than of that count alone. As in Section~\ref{sec:depth}: against timing error, pulse count is cost, every inter-block pulse leaks, and the thresholds of Table~\ref{tab:thr} should be read with an uncertainty attached to the compilation: a factor of about two for the $+\varepsilon$ thresholds of the even-response circuits, and up to a factor eight for the circuits with a first-order term and for the offset model (QPE at $\phi=0.3$ with the offset: $6.4\times10^{-2}$ main, $7.6\times10^{-3}$ LCCZ). The compilation can also change the character of the response: with the Chadwick CZ, Grover acquires an odd part with $x_c=4.6\times10^{-3}$, comparable to its threshold, and with the leakage-controlled CZ $x_c=2.8\times10^{-3}$.

\section{Decoherence and the choice of $J$}
\label{sec:deco}

Timing error is not the only thing that scales with the pulse duration. The
dominant decoherence of EO qubits in silicon is the quasi-static hyperfine
gradient~\cite{andrews2019,weinstein2023}, which we model as a static
detuning $h_i$ of every spin, drawn from $\mathcal N(0,\sigma_h)$ in units of
$J$ and applied as $\pi\sum_i h_i S_z^i\,t$ during every pulse of duration $t$
(evolved exactly together with the exchange of the pulsed pair). A gradient both dephases and leaks, because
$\sum_i h_iS^i_z$ does not conserve the block spin. For a singlet the Gaussian
$1/e$ time is $T_2^*=\tsw/(\pi\sigma_h)$, so $\sigma_h$ is the inverse of
$T_2^*$ measured in SWAP times. Table~\ref{tab:deco} (top) scans $\sigma_h$
for the three cheap circuits, 40 draws each.

\begin{table*}[t]
\centering\small
\caption{Top: quasi-static Zeeman gradient $\sigma_h$ (units of $J$; $T_2^*/\tsw =1/\pi\sigma_h$), no timing error, main compilation (the longer fixed sequences give $2$--$2.5$ times larger $\Hz$ errors at the same $\sigma_h$). Middle: $J$ trade-off at $T_2^*=3.5\,\mu$s and an uncompensated duration grid of $0.5$\,ns. Bottom: the same with continuous durations (no grid), the idealisation of amplitude control. Middle and bottom entries are ranges over three compilations (search-based single-qubit synthesis on two machines, and the fixed sequences of Section~\ref{sec:variants}); 40 draws; $\Delta E$ in mHa.}
\label{tab:deco}
\begin{tabular}{lccccc}
\toprule
$\sigma_h$ & $10^{-4}$ & $3\times10^{-4}$ & $10^{-3}$ & $3\times10^{-3}$ & $10^{-2}$\\
$T_2^*/\tsw$ & 3183 & 1061 & 318 & 106 & 32\\
\midrule
$\Hz$ $\Delta E$ [mHa] & 0.012 & 0.11 & 1.2 & 11 & 110\\
Grover $P$ (leak) & 0.933 (0.01) & 0.841 (0.10) & 0.388 (0.45) & 0.083 (0.67) & 0.022 (0.87)\\
QPE $P$ & 0.992 & 0.930 & 0.503 & 0.102 & 0.019\\
\bottomrule
\end{tabular}

\medskip
{\setlength{\tabcolsep}{3pt}\footnotesize
\begin{tabular}{lccccccc}
\toprule
$J/h$ (MHz) & 10 & 20 & 50 & 100 & 200 & 500 & 1000\\
$\tsw$ (ns) & 50 & 25 & 10 & 5 & 2.5 & 1 & 0.5\\
$\sigma_h$ & $4.5\times10^{-3}$ & $2.3\times10^{-3}$ & $9.1\times10^{-4}$ & $4.5\times10^{-4}$ & $2.3\times10^{-4}$ & $9.1\times10^{-5}$ & $4.5\times10^{-5}$\\
grid$/\tsw$ & 0.01 & 0.02 & 0.05 & 0.1 & 0.2 & 0.5 & 1\\
\midrule
$\Hz$ $\Delta E$ [mHa] & 25--58 & 7.0--17 & 2.3--26 & 14--51 & 258--344 & 92--693 & 123--259 \\
Grover $P$ & 0.044--0.049 & 0.091--0.121 & 0.345--0.417 & 0.459--0.679 & 0.017--0.033 & 0.019--0.255 & 0.000--0.141 \\
QPE $P$ & 0.033--0.064 & 0.090--0.135 & 0.356--0.531 & 0.619--0.762 & 0.176--0.442 & 0.009--0.094 & 0.000--0.562 \\
\midrule
\multicolumn{8}{l}{continuous durations, no grid:}\\
$\Hz$ $\Delta E$ [mHa] & 25--58 & 6.4--16 & 1.0--2.5 & 0.26--0.63 & 0.064--0.16 & 0.01--0.025 & 0.0026--0.0063 \\
Grover $P$ & 0.044--0.048 & 0.095--0.119 & 0.387--0.431 & 0.699--0.733 & 0.870--0.894 & 0.932--0.938 & 0.942--0.944 \\
QPE $P$ & 0.032--0.064 & 0.091--0.134 & 0.369--0.557 & 0.741--0.850 & 0.924--0.959 & 0.987--0.993 & 0.997--0.998 \\
\bottomrule
\end{tabular}}
\end{table*}

The gradient is the more demanding of the two per unit of SWAP time: a 1216-pulse circuit of about a thousand SWAP times loses $0.05$ of its success at $T_2^*/\tsw\approx1500$ (interpolated) and half at about $400$, whereas the same loss from timing needs $\varepsilon\approx6\times10^{-3}$. The reason is that the
gradient acts during every pulse of every block and leaks ($0.45$ at $\sigma_h=10^{-3}$), so its effect grows with the total duration rather than
with the error per pulse.

This makes the choice of $J$ a trade-off that the timing study alone does not
see. We fix the measured $T_2^*=3.5\,\mu$s~\cite{weinstein2023} and a
waveform-generator grid of $0.5$\,ns, and vary $J/h$ from $10$\,MHz to $1$\,GHz;
$\tsw=h/2J$ then sets both $\sigma_h=\tsw/\pi T_2^*$ and the grid in units of
$\tsw$ (Table~\ref{tab:deco}, middle). Below about $160$\,MHz the Grover circuit ($1.1\times10^3$ SWAP times) lasts longer than $T_2^*$, and below $50$\,MHz all three circuits fail by dephasing; above $100$\,MHz the grid is coarser than a tenth of $\tsw$ and they fail by quantisation; the best uncompensated operating point, near $100$\,MHz, leaves Grover at $0.46$--$0.68$ depending on the compilation and $\Hz$ outside chemical accuracy in every compilation. If instead the area can be set continuously (Table~\ref{tab:deco}, bottom) the dependence is monotonic. We simulate this as continuous durations at fixed $J$; fixed-duration amplitude control sets the same areas but exposes every pulse to the gradient for the same time, which changes the dephasing and was not simulated, so these rows are a continuous-duration idealisation rather than an amplitude-control simulation. In the weak-gradient regime every doubling of $J$ halves the gradient in units of $J$ and roughly quarters its damage; $\Hz$ regains chemical accuracy at $50$\,MHz ($100$\,MHz with the longer fixed sequences), and Grover reaches $0.70$--$0.73$ at $100$\,MHz, $0.93$--$0.94$ at $500$\,MHz and $0.94$ at $1$\,GHz in all three compilations. The non-monotonic middle rows are a commensurability effect: on the $0.5\,\tsw$ grid the $\sqrt{\mathrm{SWAP}}$ and SWAP pulses are again representable and only the single-qubit rotations are destroyed, while on the $1\,\tsw$ grid the $\sqrt{\mathrm{SWAP}}$ pulses ($\tfrac12$, $\tfrac32$) are rounded to $0$ and $2$, i.e.\ to the identity. Within this model, higher $J$ is better as long as the area can be set continuously; lowering $J$ to make a duration grid finer buys timing precision at a higher cost in dephasing. Weinstein \emph{et al.} note that the flat-barrier regime above $20$\,GHz, where charge noise is smallest, is too fast for their pulsed operation; the trade-off of Table~\ref{tab:deco} is then between the few hundred MHz that pulse shaping permits and the $T_2^*$ of the device.

\section{What the thresholds mean for hardware}
\label{sec:hardware}

Table~\ref{tab:thr} puts the end of the useful regime at a fractional error of
$3$--$9\times10^{-3}$ for circuits whose odd part is negligible at the
threshold (VQE, QAOA at $p=2$, exact QPE, Grover), $3$--$8\times10^{-3}$ for jitter on the larger circuits, and $7\times10^{-3}$ to $1.1\times10^{-2}$
(systematic) or $5\times10^{-3}$ (jitter) for the repetition code's logical
flip; Section~\ref{sec:variants} attaches a factor of two to five to these numbers.
Above $x_c$ precision enters squared: a tenfold better area precision is a
hundredfold smaller error. The question is what ``$10^{-3}$ of the SWAP
time'' means for a real control chain.

\emph{Duration grids.} With $J/h$ between $10$\,MHz and a few hundred MHz~\cite{andrews2019,weinstein2023} the SWAP time is $1$--$50$\,ns.
We simulated a duration grid directly: every nominal duration rounded to the
nearest multiple of $\Delta t$, no amplitude compensation
(Table~\ref{tab:grid}). The entangling sequence uses only multiples of
$\tsw/2$, so it is exact on every grid that divides $\tsw/2$ (all tested grids
up to $0.1\,\tsw$); what such a grid spoils is the single-qubit rotations. Because the rounding depends on which single-qubit decomposition is chosen, Table~\ref{tab:grid} gives, for each grid, the range over three compilations (the search-based synthesis on two machines, which select different but equivalent four-pulse solutions, and the fixed sequences of Section~\ref{sec:variants}). A grid of $0.02\,\tsw$ keeps $\Hz$ within chemical accuracy and Grover above $0.93$ in all three; at $0.05\,\tsw$ $\Hz$ loses chemical accuracy in two of the three and Grover drops to $0.85$--$0.93$; at $0.1\,\tsw$ Grover ranges from $0.53$ to $0.93$, the largest compilation dependence in this study; on the $0.2\,\tsw$ grid the $\sqrt{\mathrm{SWAP}}$ pulses themselves are rounded, the entangler leaks ($0.67$ for Grover in the main compilation) and every circuit fails. The boundary therefore lies between $2\%$ and $5\%$ of $\tsw$, i.e.\ at $0.1$--$0.25$\,ns for a $5$\,ns SWAP, with the exact position depending on the compilation. There are two ways out. Chadwick \emph{et al.}~\cite{chadwick2024} assume that every pulse has the same duration and the area is set by the amplitude. The HRL cryo-CMOS controller~\cite{hrl2026} does the opposite: it switches between fixed, calibrated DAC voltages and sets each exchange angle by a pulse duration of $0.4$--$6$\,ns from a digital-to-time converter with a step of about $1$\,ps, i.e.\ $\Delta t/\tsw\approx10^{-3}$ or finer, far below the $2\%$ at which Table~\ref{tab:grid} starts to show damage; its pulse-width jitter, $4$--$7.5\times10^{-4}$ of the width, is the $\sigma_{\mathrm m}$ of our model at a level that costs the circuits of this study less than $0.01$ in success probability (Table~\ref{tab:thr}: $\sigma_{\mathrm m}$ thresholds $4$--$8\times10^{-3}$). Either way the timing precision of this paper is the precision of the exchange area, and $\varepsilon$ is its fractional error.

\begin{table}[t]
\centering\small
\caption{Duration grid without amplitude compensation: every nominal duration
rounded to a multiple of $\Delta t$ (units of $\tsw$); ranges over three
compilations (search-based synthesis on two machines, fixed sequences). The
entangler is exact on all grids up to $0.1$; on $0.2$ its
$\sqrt{\mathrm{SWAP}}$ pulses are rounded and it leaks.}
\label{tab:grid}\label{tab:gridrange}
\setlength{\tabcolsep}{3pt}
\resizebox{\columnwidth}{!}{\begin{tabular}{lccc}
\toprule
$\Delta t/\tsw$ & $\Hz$ $\Delta E$ [Ha] & Grover $P$ & QPE $P$ ($\phi=3/8$)\\
\midrule
0.005 & $2.5\times10^{-5}$--$\,3.3\times10^{-5}$ & 0.938--0.944 & 1.000--1.000 \\
0.01 & $5.9\times10^{-5}$--$\,1.8\times10^{-4}$ & 0.943--0.950 & 0.998--0.999 \\
0.02 & $3.3\times10^{-4}$--$\,5.0\times10^{-4}$ & 0.937--0.952 & 0.988--0.994 \\
0.05 & $1.4\times10^{-3}$--$\,2.3\times10^{-2}$ & 0.846--0.928 & 0.903--0.939 \\
0.1 & $1.4\times10^{-2}$--$\,5.1\times10^{-2}$ & 0.535--0.934 & 0.805--0.885 \\
0.2 & $2.6\times10^{-1}$--$\,3.4\times10^{-1}$ & 0.016--0.029 & 0.193--0.457 \\
\bottomrule
\end{tabular}}
\end{table}

\emph{Jitter at realistic levels.} A clock jitter of $3$\,ps on a $10$--$50$\,ns pulse is $\sigma_{\mathrm a}\approx10^{-4}$, and because it
accumulates linearly it costs the 1216-pulse Grover circuit a negligible amount ($0.945\to0.943$ at $\sigma_{\mathrm a}=10^{-3}$).

\emph{Calibration drift and charge noise.} A fractional error
of the exchange area that is constant over one shot --- a miscalibrated
$J(V)$, or the part of charge noise that is quasi-static over the
$5$--$50\,\mu$s of a thousand pulses --- behaves as $\varepsilon$: it multiplies
every pulse of the run, accumulates as $N^2$, and whether its sign changes
from shot to shot does not matter for an even response. Such an error is not
the same on every bond, so we also drew an independent static fractional error
for each bond (standard deviation $s$, 20 draws). In this model it is two to
three times as damaging as a common $\varepsilon=s$: at $s=10^{-2}$ the $\Hz$
energy error is $1.2\times10^{-2}$\,Ha against $4.7\times10^{-3}$, Grover's
success $0.67$ against $0.77$, exact QPE's $0.87$ against $0.94$; at
$s=3\times10^{-3}$ the $\Hz$ error is $1.1\times10^{-3}$\,Ha, close to but
within chemical accuracy. Part of the difference is the loss of the SWAP's
common-mode immunity.

\emph{Two measurements fix the scale.} Andrews \emph{et
al.}~\cite{andrews2019} deliberately scaled every exchange angle of their
single-qubit Cliffords by $(1\pm\Delta)$ with random signs and measured the
error per Clifford by blind randomized benchmarking: the added error grows as $4.3\,\Delta^2$ (close to the expected $\pi^2/6$ per rotation times $2.7$ rotations per Clifford) on top of a $0.42\%$ offset, reaches $2.0\%$ at $\Delta=6\%$, and the leakage stays at $0.15\%$ throughout.
This is the behaviour of our single-qubit gates, which do not leak and whose even coefficients are $1.4$ ($X$) and $1.2$ ($H$) per four-pulse gate. From the same data they infer that the combined effect of imperfect control and charge noise in their device is equivalent to a random over-rotation of about $1.3\%$ rms. A random sign per angle is not a common-mode $\varepsilon$; in our models it is closest to the bond-dependent static error $s$ (which costs two to three times more than a common $\varepsilon$ of the same size) or to multiplicative jitter (which costs less than a common $\varepsilon$ of the same size for the shallow circuits and about the same for Grover). Weinstein \emph{et al.}~\cite{weinstein2023} characterise charge noise
by the number $N_{\rm osc}$ of coherent exchange oscillations before the
Gaussian $1/e$ decay: about $50$ at $J/h\approx100$\,MHz, below $100$ at
$1$\,GHz, above $1600$ when the barrier flattens above $20$\,GHz. For
quasi-static Gaussian fluctuations of $J$, $N_{\rm osc}=\sqrt2\,J/2\pi\sigma_J$,
so $\sigma_J/J\approx0.23/N_{\rm osc}$: $4.5\times10^{-3}$ at $100$\,MHz and below $1.4\times10^{-4}$ in the flat-barrier regime. Andrews \emph{et al.}'s $1.3\%$ and Weinstein \emph{et al.}'s $4.5\times10^{-3}$ at $100$\,MHz place today's hardware at the edge of Table~\ref{tab:thr}, while the flat-barrier value would lie well inside it: at an area error of $5\times10^{-3}$--$10^{-2}$ VQE sits around chemical accuracy, Grover near $0.7$--$0.9$ instead of $0.95$, the two-qubit gate at $10^{-3}$--$5\times10^{-3}$ infidelity with a third of it leakage. The same device's single-qubit error of $1.1\times10^{-3}$ with leakage $3\times10^{-4}$ and its Fong--Wandzura CNOT fidelity of $96.3\%$~\cite{weinstein2023} are consistent with an area precision in that range plus the magnetic decoherence of Section~\ref{sec:deco}; randomized-benchmarking errors average over the Clifford group and are not the coherent gate infidelities of Table~\ref{tab:gates}, so the comparison is one of scale, not of value. An order of magnitude in area stability moves every even-response circuit of this study two orders of magnitude inside its useful regime.

\emph{The bit budget.} Table~\ref{tab:thr} can be restated in the units of a
digital control system. A control word with $b$ significant bits sets the pulse area, or at fixed $J$ the duration (the two are the same fractional error), with a relative rounding error of at most $2^{-b}$. Read as a common-mode calibration error of that size --- the one direction in the space of per-pulse errors that we scanned, which does not bound errors with different signs on different pulses --- the thresholds of $3$--$9\times10^{-3}$ correspond to $b=7$--$9$, and for an even response one more bit reduces the damage by a factor four. Thus $8$--$10$ significant bits of common-mode-equivalent precision put every circuit of this study inside its useful regime. The $11$-bit significand of IEEE half precision ($10$ stored bits; rounding $\le4.9\times10^{-4}$) leaves the damage of the even-response circuits $40$--$300$ times below their thresholds (QPE at $\phi=0.3$, with its first-order term, only eight times), while the $8$-bit significand of bfloat16 ($7$ stored bits; rounding $\le3.9\times10^{-3}$) sits at the threshold of the most sensitive circuits and inside for the rest. For
the duration we simulated the rounding directly: Table~\ref{tab:fp} rounds
every nominal duration to a binary mantissa of $b$ bits (a floating-point
rather than fixed-point grid, so short pulses keep a fine absolute step). The requirement found this way is lower than the common-mode figure, and it depends on the compilation, so Table~\ref{tab:fp} gives ranges over the same three compilations as Table~\ref{tab:grid}: $5$ bits keep Grover and QPE within their $5\%$ in all three, $6$ bits keep $\Hz$ within chemical accuracy and the CNOT below $10^{-3}$ in all three, and $7$ bits leave $\Hz$ below $4\times10^{-4}$\,Ha, the losses of Grover and QPE below $0.003$ (one compilation is $0.005$ above ideal) and the CNOT below $10^{-4}$ in all three. The main reason is that the entangling sequence uses only the dyadic durations $\tfrac12$, $1$, $\tfrac32$, which are exact in any binary word of two or more bits, so that rounding touches only the single-qubit rotations. These statements hold for the tested circuits and compilations; they are not a bound for arbitrary error patterns. This protection is specific to sequences with dyadic angles (the Zeuch--Bonesteel sequence, the Chadwick CZ and the logical SWAP, not the leakage-controlled CZ, five of whose angles are not dyadic) and was tested only for the three circuits of the table. Clock jitter at the level of the controller discussed above ($\le10^{-3}$) costs these circuits little relative to the budget (Grover $0.945\to0.943$ at $\sigma_{\mathrm a}=10^{-3}$; Table~\ref{tab:thr}). The quantity that the devices of Refs.~\cite{andrews2019,weinstein2023} lacked is therefore not bits: with the proxy $b_{\mathrm{eff}}=-\log_2\sigma$ for an rms fractional area error $\sigma$, Andrews \emph{et al.}'s $1.3\%$ equivalent over-rotation and Weinstein \emph{et al.}'s $4.5\times10^{-3}$ at $100$\,MHz are $6$--$8$ effective bits of analog area stability, against the $8$--$10$ the algorithms need; a DAC's nominal $14$--$16$ bits cannot be compared with this without the voltage range and the slope of $J(V)$ at the operating point. Van Dijk \emph{et al.}~\cite{vandijk2019}
reached the same kind of number from the gate side for single-spin qubits
($8$ bits of arbitrary-waveform-generator (AWG) resolution for a $99.9\%$ $\pi$ rotation, $10$ for the detuning of a two-qubit gate and $9$ for read-out); Bhattacharjee \emph{et al.}~\cite{bhattacharjee2026} find that a quantum neural network keeps its accuracy with $6$-bit DACs at inference ($>90\%$ with $4$--$5$ bits). Our numbers sit between the two, as expected for a different figure of merit on a different platform; in all three cases the resolution requirement is modest compared with available hardware, and for the EO analysis here the limiting quantity is stability.

\begin{table*}[t]
\centering\small
\caption{Every nominal pulse duration rounded to a binary floating-point significand of $b$ bits including the leading bit (relative step $2^{-(b-1)}$), no other error; ranges over three compilations (search-based synthesis on two machines, fixed sequences). The entangling sequence's durations are dyadic and survive any $b\ge2$.}
\label{tab:fp}
\begin{tabular}{rccccc}
\toprule
$b$ & step & $\Hz$ $\Delta E$ [Ha] & Grover $P$ & QPE $P$ & CNOT infid.\\
\midrule
4 & $1.2\times10^{-1}$ & $8.1\times10^{-3}$--$\,1.6\times10^{-2}$ & 0.901--0.950 & 0.917--0.951 & $8.1\times10^{-3}$--$\,1.1\times10^{-2}$ \\
5 & $6.2\times10^{-2}$ & $6.3\times10^{-4}$--$\,2.5\times10^{-3}$ & 0.922--0.947 & 0.981--0.989 & $1.3\times10^{-3}$--$\,2.3\times10^{-3}$ \\
6 & $3.1\times10^{-2}$ & $3.1\times10^{-4}$--$\,1.2\times10^{-3}$ & 0.917--0.958 & 0.995--0.995 & $3.5\times10^{-4}$--$\,8.1\times10^{-4}$ \\
7 & $1.6\times10^{-2}$ & $2.8\times10^{-5}$--$\,3.5\times10^{-4}$ & 0.943--0.951 & 0.998--0.999 & $2.4\times10^{-5}$--$\,7.4\times10^{-5}$ \\
8 & $7.8\times10^{-3}$ & $2.3\times10^{-5}$--$\,5.2\times10^{-5}$ & 0.943--0.944 & 1.000--1.000 & $4.8\times10^{-6}$--$\,2.1\times10^{-5}$ \\
11 & $9.8\times10^{-4}$ & $8.0\times10^{-8}$--$\,9.8\times10^{-7}$ & 0.945--0.945 & 1.000--1.000 & $1.1\times10^{-7}$--$\,7.7\times10^{-7}$ \\
16 & $3.1\times10^{-5}$ & $7.1\times10^{-10}$--$\,9.4\times10^{-10}$ & 0.945--0.945 & 1.000--1.000 & $6.1\times10^{-10}$--$\,1.2\times10^{-9}$ \\
\bottomrule
\end{tabular}
\end{table*}

\emph{The HRL controller as a check.} The cryo-controlled device of Ref.~\cite{hrl2026} reports mean single-qubit gate errors of $2\times10^{-4}$ and CNOT errors of $3\times10^{-3}$ (best $9\times10^{-4}$) from randomized benchmarking, a median $N_{\rm osc}$ of $674$ and $T_2^*=19.3\,\mu$s measured on a similar device, and a distance-3 repetition code with logical error $3.2\times10^{-3}$ per round once leakage-reduction units are included. Read with our CNOT scan ($53\,\varepsilon^2$), a CNOT error of $3\times10^{-3}$ would be an area error of $7.5\times10^{-3}$ and $9\times10^{-4}$ one of $4\times10^{-3}$, i.e.\ $7$--$8$ effective bits, whereas the device's exchange noise alone, $\sigma_J/J\approx3.3\times10^{-4}$ by the conversion above, would be $11.5$. On that device the gate error is not set by the exchange noise but, as the authors state, by static gradients and contextual miscalibration, which our gradient model and $\varepsilon$ represent in kind if not in number. The comparison is one of scale: their 37-pulse CNOT and benchmarking average are not our 28-pulse coherent infidelity. That their repetition code needs leakage-reduction units is what the one-round result of Section~\ref{sec:algos} would predict for a decoder blind to leakage; we have not simulated the units themselves.

\emph{Two possible remedies.} Dynamically corrected exchange sequences~\cite{hickman2013} cancel the leading-order effect of exchange and hyperfine-gradient errors together at the cost of longer sequences; their effect at the algorithm level, against an $N^2$ accumulation on one side and a longer exposure to the gradient on the other, is not assessed here.
And the common-mode immunity of the logical SWAP shows what immunity looks
like at the sequence level: a first-order error operator that is a function
of quantities conserved on the code space. Whether an entangling sequence
with that property exists is not settled here.

\section{What this study does not establish}
\label{sec:limits}

Apart from the timing error and, in Section~\ref{sec:deco}, a quasi-static
gradient, the simulation is noiseless: no charge-noise spectrum beyond its
quasi-static part, no state-preparation or measurement error, no shot noise,
no crosstalk between simultaneously pulsed pairs, no finite rise time (an
offset $\delta$ stands in for it). The circuits are small ($\le6$ encoded
qubits, $\le2170$ pulses). Three entangling sequences and two single-qubit
compilations were tried; the compilation moves the thresholds by a factor of two to eight and we did not search for the most robust one; LiH and QAOA were not repeated with the Chadwick entanglers. The QAOA angles and VQE
parameters were optimised on the ideal device and re-optimised only locally,
without monitoring convergence. The chemical-accuracy threshold is applied to
the active-space energy. The VQE energy is the measured estimator of the
device, not the energy of one prepared state (Table~\ref{tab:decomp} is the
only place where the two are separated). Success probabilities for Grover, QPE and the QAOA optimal cut assume leakage detection; the VQE energies, the QAOA ratios and the repetition code use the physical readout, and the code was run for one round with ideal decoding.
The mapping of $\varepsilon$ to charge noise assumes the noise is quasi-static
over a run, the bond-dependent model assumes independent bonds, and the
gradient model assumes a static, Gaussian, independent field per spin; all
three are statements about the noise spectrum of a specific device. The
literature values quoted in Section~\ref{sec:hardware} are read from the published figures and text of Refs.~\cite{andrews2019,weinstein2023} and converted with a quasi-static assumption that those authors do not make. Finally, the selection of the shortest among several numerically equivalent single-qubit decompositions depends on floating-point details, so results are reproducible on one machine but not bit-for-bit between machines: for the continuous-duration and gradient scans two machines agreed within $5\%$ in the $\Hz$ energy error and within $16\%$ in the Grover and QPE success probabilities, but for the duration-grid and rounding scans, where the result depends on which durations happen to fall near a rounding boundary, individual entries differed by up to a factor of seven to eight, and qualitatively for entries near zero (Tables~\ref{tab:gridrange}, \ref{tab:fp} and the middle and bottom of Table~\ref{tab:deco} therefore give ranges). All other numbers were produced on one machine.

\section{Conclusion}

A single number, the fractional precision of the exchange area, governs how
far an exchange-only processor can take an algorithm whose ideal answer is an
extremum of its figure of merit: across VQE, QAOA, Grover, QPE and a small
code the useful regime ends at a few $10^{-3}$, with a weak dependence on the algorithm and the depth and a dependence of a factor two to five on the compilation. A
third of the CNOT error is leakage from pulses between blocks; it is not
absorbed by variational re-optimisation, is not reduced by a leakage-controlled entangler, and makes a bit-flip code that is blind to it correct in the wrong direction, by a factor $38$--$93$ at $\varepsilon=10^{-2}$ in three of four compilations; in the fourth the uncorrected data already fail at the same rate. Systematic errors accumulate as the square of the depth
and bias the answer; jitter accumulates linearly and leaves the mean of phase estimation unchanged within statistics. Measured over-rotation and charge noise put current silicon devices
at the edge of this regime, and a quasi-static magnetic gradient with the
measured $T_2^*$ limits thousand-pulse circuits unless $J$ is run at several hundred MHz with the area set continuously, by amplitude or by a picosecond timer, because a duration grid of $2$--$5\%$ of the SWAP time is where quantisation starts to cost chemical accuracy. In the units of a control system a common-mode area error needs $8$--$10$ significant bits and, in the tested circuits, a $6$-bit significand for the durations suffices; half-precision words and picosecond timers already exceed this, and the bits that older devices lacked were analog ones.

\section*{Code and data}

The simulator, the compiler, the experiment scripts and all numerical results
(JSON, with the time of each measurement) are at
\url{https://github.com/minatoyuichiro/eotime}; the tables marked as generated are produced from the JSON files by the scripts, and the hand-typed tables are checked against them.

\bibliographystyle{apsrev4-2}
\bibliography{refs}

@article{bacon2000,
  author  = {Bacon, Dave and Kempe, Julia and Lidar, Daniel A. and Whaley, K. Birgitta},
  title   = {Universal Fault-Tolerant Quantum Computation on Decoherence-Free Subspaces},
  journal = {Physical Review Letters},
  volume  = {85},
  pages   = {1758--1761},
  year    = {2000},
  doi     = {10.1103/PhysRevLett.85.1758}
}

@article{divincenzo2000,
  author  = {DiVincenzo, D. P. and Bacon, D. and Kempe, J. and Burkard, G. and Whaley, K. B.},
  title   = {Universal quantum computation with the exchange interaction},
  journal = {Nature},
  volume  = {408},
  pages   = {339--342},
  year    = {2000},
  doi     = {10.1038/35042541}
}

@article{kempe2001,
  author  = {Kempe, J. and Bacon, D. and Lidar, D. A. and Whaley, K. B.},
  title   = {Theory of decoherence-free fault-tolerant universal quantum computation},
  journal = {Physical Review A},
  volume  = {63},
  pages   = {042307},
  year    = {2001},
  doi     = {10.1103/PhysRevA.63.042307}
}

@article{fong2011,
  author  = {Fong, Bryan H. and Wandzura, Stephen M.},
  title   = {Universal quantum computation and leakage reduction in the 3-qubit decoherence free subsystem},
  journal = {Quantum Information and Computation},
  volume  = {11},
  pages   = {1003--1018},
  year    = {2011},
  note    = {arXiv:1102.2909}
}

@article{zeuch2016,
  author  = {Zeuch, Daniel and Bonesteel, N. E.},
  title   = {Simple derivation of the {Fong--Wandzura} pulse sequence},
  journal = {Physical Review A},
  volume  = {93},
  pages   = {010303(R)},
  year    = {2016},
  doi     = {10.1103/PhysRevA.93.010303},
  note    = {arXiv:1508.07524}
}

@article{setiawan2014,
  author  = {Setiawan, F. and Hui, Hoi-Yin and Kestner, J. P. and Wang, Xin and Das Sarma, S.},
  title   = {Robust two-qubit gates for exchange-coupled qubits},
  journal = {Physical Review B},
  volume  = {89},
  pages   = {085314},
  year    = {2014},
  doi     = {10.1103/PhysRevB.89.085314}
}

@article{chadwick2024,
  author  = {Chadwick, Jason D. and Guerreschi, Gian Giacomo and Luthi, Florian and M{\k{a}}dzik, Mateusz T. and Mohiyaddin, Fahd A. and Prabhu, Prithviraj and Schmitz, Albert T. and Litteken, Andrew and Premaratne, Shavindra and Bishop, Nathaniel C. and Matsuura, Anne Y. and Clarke, James S.},
  title   = {Short two-qubit pulse sequences for exchange-only spin qubits in two-dimensional layouts},
  journal = {Physical Review A},
  volume  = {111},
  pages   = {052616},
  year    = {2025},
  doi     = {10.1103/PhysRevA.111.052616},
  note    = {arXiv:2412.14918}
}

@misc{chadwickdata,
  author = {Chadwick, Jason D. and others},
  title  = {Data for ``{Short} two-qubit pulse sequences for exchange-only spin qubits in {2D} layouts'', pulse-sequence library v1.6.0},
  howpublished = {Zenodo, \url{https://doi.org/10.5281/zenodo.14518251}},
  year   = {2024},
  doi    = {10.5281/zenodo.14518251}
}

@article{andrews2019,
  author  = {Andrews, R. W. and Jones, C. and Reed, M. D. and Jones, A. M. and Ha, S. D. and Jura, M. P. and Kerckhoff, J. and Levendorf, M. and Meenehan, S. and Merkel, S. T. and Smith, A. and Sun, B. and Weinstein, A. J. and Rakher, M. T. and Ladd, T. D. and Borselli, M. G.},
  title   = {Quantifying error and leakage in an encoded {Si/SiGe} triple-dot qubit},
  journal = {Nature Nanotechnology},
  volume  = {14},
  pages   = {747--750},
  year    = {2019},
  doi     = {10.1038/s41565-019-0500-4}
}

@article{weinstein2023,
  author  = {Weinstein, Aaron J. and Reed, Matthew D. and Jones, Aaron M. and Andrews, Reed W. and Barnes, David and Blumoff, Jacob Z. and Euliss, Larken E. and Eng, Kevin and Fong, Bryan H. and Ha, Sieu D. and Hulbert, Daniel R. and Jackson, Clayton A. W. and Jura, Michael and Keating, Tyler E. and Kerckhoff, Joseph and Kiselev, Andrey A. and Matten, Justine and Sabbagh, Golam and Smith, Aaron and Sun, Bo and Weigand, Alexander J. and Wright, Jeffrey and Rakher, Matthew T. and Ladd, Thaddeus D. and Borselli, Matthew G.},
  title   = {Universal logic with encoded spin qubits in silicon},
  journal = {Nature},
  volume  = {615},
  pages   = {817--822},
  year    = {2023},
  doi     = {10.1038/s41586-023-05777-3}
}

@article{hickman2013,
  author  = {Hickman, G. T. and Wang, Xin and Kestner, J. P. and Das Sarma, S.},
  title   = {Dynamically corrected gates for an exchange-only qubit},
  journal = {Physical Review B},
  volume  = {88},
  pages   = {161303(R)},
  year    = {2013},
  doi     = {10.1103/PhysRevB.88.161303}
}

@misc{epo,
  title  = {exchange-pulse-optimizer: placement, routing and scheduling for exchange-only encoded qubits},
  howpublished = {\url{https://github.com/kaluza1/exchange-pulse-optimizer}},
  author = {{kaluza1 (GitHub user)}},
  note   = {{GitHub} repository, no author name or citation file given; accessed October 2026},
  year   = {2025}
}

@article{openfermion,
  author  = {McClean, Jarrod R. and others},
  title   = {{OpenFermion}: the electronic structure package for quantum computers},
  journal = {Quantum Science and Technology},
  volume  = {5},
  pages   = {034014},
  year    = {2020},
  doi     = {10.1088/2058-9565/ab8ebc}
}

@misc{qiskitH2,
  author = {{Qiskit community}},
  title  = {Simulating molecules using {VQE} ({Qiskit} textbook): two-qubit parity-mapped {H$_2$} {Hamiltonian} at 0.735~{{\AA}}},
  howpublished = {\url{https://github.com/Qiskit/textbook} (archived textbook)},
  year   = {2021}
}

@article{hrl2026,
  author  = {{HRL Quantum Team} and Abraham, Michael and Acuna, Edwin and others},
  title   = {A digitally controlled silicon quantum processing unit},
  journal = {Nature},
  volume  = {655},
  pages   = {1154--1159},
  year    = {2026},
  doi     = {10.1038/s41586-026-10754-7},
  note    = {arXiv:2604.16216}
}

@article{vandijk2019,
  author  = {van Dijk, Jeroen P. G. and Kawakami, Erika and Schouten, Raymond N. and Veldhorst, Menno and Vandersypen, Lieven M. K. and Babaie, Masoud and Charbon, Edoardo and Sebastiano, Fabio},
  title   = {Impact of classical control electronics on qubit fidelity},
  journal = {Physical Review Applied},
  volume  = {12},
  pages   = {044054},
  year    = {2019},
  doi     = {10.1103/PhysRevApplied.12.044054}
}

@misc{bhattacharjee2026,
  author = {Bhattacharjee, Rupayan and Sarkar, Rohit Sarma and Abadal, Sergi and Almud{\'e}ver, Carmen G. and Alarc{\'o}n, Eduard},
  title  = {Assessing the impact of low resolution control electronics on quantum neural network performance},
  year   = {2026},
  note   = {arXiv:2601.04983}
}

\end{document}